\documentclass[twocolumn]{article}
\usepackage{amssymb}

\usepackage{graphicx}
\usepackage{amsmath}

\begin{document}

\title{Crowd-Anticrowd Theory of Multi-Agent Market Games}
\author{M. Hart$^{1}$, P. Jefferies$^{1}$, P.M. Hui$^{2}$ and N.F. Johnson$^{1}$ \\
$^{1}$Physics Department, Oxford University, Oxford, OX1 3PU, U.K.\\
$^{2}$Physics Department, Chinese University of Hong Kong, Shatin, Hong Kong}
\maketitle

\begin{abstract}
We present a dynamical theory of a multi-agent market game, the so-called
Minority Game (MG), based on crowds and anticrowds. The time-averaged
version of the dynamical equations provides a quantitatively accurate, yet
intuitively simple, explanation for the variation of the standard deviation
(`volatility') in MG-like games. We demonstrate this for the basic MG, and
the MG with stochastic strategies. The time-dependent equations themselves
reproduce the essential dynamics of the MG.
\end{abstract}

\bigskip \bigskip

Agent-based games have great potential application in the study of
fluctuations in financial markets. Challet and Zhang's Minority Game (MG) 
\cite{challet,econophysics} offers possibly the simplest example and has
been the subject of much research \cite{econophysics}. The MG comprises an
odd number of agents $N$ choosing repeatedly between option 0 (e.g. buy) and
option 1 (e.g. sell). The winners are those in the minority group, e.g.
sellers win if there is an excess of buyers. The outcome at each timestep
represents the winning decision, 0 or 1. A common bit-string of the $m$ most
recent outcomes is made available to the agents at each time-step \cite
{memory}. The agents randomly pick $s$ strategies at the beginning of the
game, with repetitions allowed - each strategy is a bit-string of length $%
2^{m}$ which predicts the next outcome for each of the $2^{m}$ possible
histories. Agents reward successful strategies with a (virtual) point. At
each turn of the basic MG, the agent uses her most successful strategy, i.e.
the one with the most virtual points. Here we develop a dynamical theory for
MG-like games based on the formation of crowds and anticrowds.

The number of agents holding a particular combination of strategies can be
written as a $D\times D \times \dots$ ($s$ terms) dimensional matrix $\Omega$%
, where $D$ is the total number of available strategies. For $s=2$, this is
simply a $D\times D$ matrix where the entry $(i,j)$ represents the number of
agents who picked strategy $i$ and then $j$. The strategy labels are given
by the decimal representation of the strategy plus unity, for example the
strategy 0101 for $m=2$ has strategy label 5+1=6. $\Omega $ is fixed at the
beginning of the game (`quenched disorder') and can represent either the
full strategy space or the reduced strategy space \cite{challet}, depending
on the choice of $D$. $\Sigma $ is another time-independent matrix,
containing all the strategies in the required space in their binary form: $%
\Sigma _{r,h+1}$ describes the prediction of strategy $r$ given the history $%
h$ (where $h$ is the decimal corresponding to the $m$-bit binary history
string).

We introduce a vector $\underline{n}(t)$: this contains the number of agents
using each strategy at time $t$, in order of increasing strategy label. The
vector \underline{$S$}$(t)$ contains the virtual score for each strategy at
time $t$ in order of increasing strategy label. The vector \underline{$R$}$%
(t)$ lists the strategy label in order of best-to-worst virtual points score
at time $t$; if any strategies are tied in points then the strategy with the
lower-value label is listed first. The vector \underline{$\rho $}$(t)$ shows
the rank of the strategy listed in order of increasing strategy label at
time $t$. Hence \underline{$R$}$(t)$ and \underline{$\rho $}$(t)$ can be
found from \underline{$S$}$(t)$ using simple sort operations. The vector $%
\underline{n}(t)$ is the sum of two terms 
\begin{equation}
\underline{n}(t)=\underline{n}^{0}(t)+\underline{n}^{d}(t)\ \ .
\end{equation}
Here $\underline{n}^{0}(t)$ gives the number of agents using each strategy;
however where any strategies are tied in virtual score, $\underline{n}^{0}(t)
$ assumes that the agent will use the strategy with the lower-value label by
virtue of the definition of \underline{$R$}$(t)$. The term $\underline{n}%
^{d}(t)$ accounts for tied strategies, and hence provides a correction to $%
\underline{n}^{0}(t)$. $\underline{n}^{0}(t)$ is given by 
\begin{equation}
\underline{n}^{0}(t)_{r}=\sum_{i=\rho (t)_{r}}^{2^{m+1}}[\widehat{\digamma }%
(\Omega )]_{r,R(t)_{i}}
\end{equation}
where $[\widehat{\digamma }(\Omega )]_{\alpha ,\beta }=\Omega _{\alpha
,\beta }+\Omega _{\beta ,\alpha }-\delta _{\alpha ,\beta }\Omega _{\alpha
,\beta }$. The vector $\underline{n}^{d}(t)$ is given by 
\begin{equation}
\underline{n}^{d}(t)_{r}=\sum_{r^{\prime }\neq r}\delta _{s(t)_{r^{\prime
}},s(t)_{r}}Sgn(r^{\prime }-r)Bin_{r^{\prime },r}
\end{equation}
where$:Bin_{r^{\prime },r}\thicksim B[(\widehat{\digamma }(\Omega
))_{r^{\prime },r},\frac{1}{2}]\;$and$\;Bin_{r^{\prime },r}=Bin_{r,r^{\prime
}.}$ The standard notation $Bin$ represents the binary distribution. Note
the condition $Bin_{r^{\prime },r}=Bin_{r,r^{\prime }}$ which guarantees
conservation of agents, as in the basic MG. The outcome parameter $\Upsilon
(t)$ denotes which choice, 0 or 1, is the minority (and hence winning)
decision at time $t$: 
\begin{equation}
\Upsilon (t)=\mathcal{H}[-[\underline{n}(t)^{T}\Sigma ^{\prime }]_{h(t)+1}]
\end{equation}
where $\Sigma ^{\prime }=2\Sigma -1$. The history, i.e. bit-string of the $m$
most recent outcomes, and the virtual scores of the strategies are updated
as follows: 
\begin{equation}
h(t+1)=2[h(t)-2^{m-1}\mathcal{H}[h(t)-2^{m-1}]]+\Upsilon (t)
\end{equation}
where $\mathcal{H}$ is the Heaviside function, and 
\begin{equation}
\underline{S}(t+1)=\underline{S}(t)+\Sigma _{h(t)+1}^{\prime }[2\Upsilon
(t)-1]\ \ .
\end{equation}
Equations (1-6) are a set of time-dependent equations which reproduce the
essential dynamics of the basic MG, and can be easily extended to describe
MG generalizations. Iterating these equations is equivalent to running a
numerical simulation, but is far easier and can even be done analytically. A
slight difference may arise as a result of the method chosen for
tie-breaking between strategies with equal virtual points: a numerical
program will typically break this tie using a separate coin-toss for each
agent, whereas the dynamical equations group together those agents using the
same pair of strategies and then assign a proportion of that group to a
particular strategy using a coin-toss. This difference is typically
unimportant.

As an example of the implementation of these equations, consider a time $%
t_{e}$ during the following game: $m=2$, $s=2$\ and $N=101$\ in the reduced
strategy space, with a strategy configuration $\Omega $\ and strategy score
given as follows:

\bigskip

$\Omega =\left( 
\begin{array}{cccccccc}
2 & 3 & 2 & 3 & 5 & 3 & 1 & 1 \\ 
1 & 3 & 2 & 2 & 2 & 1 & 2 & 1 \\ 
1 & 0 & 2 & 0 & 1 & 3 & 1 & 3 \\ 
1 & 1 & 0 & 1 & 1 & 0 & 1 & 3 \\ 
4 & 5 & 1 & 1 & 2 & 0 & 0 & 0 \\ 
2 & 1 & 2 & 1 & 0 & 2 & 0 & 4 \\ 
1 & 2 & 1 & 2 & 0 & 0 & 2 & 4 \\ 
1 & 2 & 2 & 1 & 1 & 1 & 1 & 2
\end{array}
\right) $

\bigskip

$\underline{S}(t_{e})=\left( 
\begin{array}{c}
3 \\ 
-1 \\ 
-3 \\ 
1 \\ 
-1 \\ 
3 \\ 
1 \\ 
-3
\end{array}
\right) ,$ with $\Sigma =\left( 
\begin{array}{cccc}
0 & 0 & 0 & 0 \\ 
0 & 0 & 1 & 1 \\ 
0 & 1 & 0 & 1 \\ 
0 & 1 & 1 & 0 \\ 
1 & 0 & 0 & 1 \\ 
1 & 0 & 1 & 0 \\ 
1 & 1 & 0 & 0 \\ 
1 & 1 & 1 & 1
\end{array}
\right) \ .$ Using these values for $\Omega $\ and $\underline{S}(t_{e})$\
we can obtain values for $\underline{n}(t)$ and ultimately $\underline{S}%
(t_{e}+1)$.\ $\Omega $\ and $\underline{S}(t_{e})$ imply that

$\underline{n}^{0}(t_{e})=\left( 
\begin{array}{c}
31 \\ 
15 \\ 
7 \\ 
13 \\ 
5 \\ 
15 \\ 
13 \\ 
2
\end{array}
\right) ,$\ and $\underline{n}^{d}(t_{e})=\left( 
\begin{array}{c}
-3 \\ 
-2 \\ 
-5 \\ 
0 \\ 
2 \\ 
3 \\ 
0 \\ 
5
\end{array}
\right) $ with probability $\frac{105}{65536}$, yielding $\underline{n}%
(t_{e})$ when summed. (When two strategies are
tied, agents holding these strategies each flip a coin to decide which
strategy to use. The separate probabilities for all tied strategies, when
multiplied together, yield the probability of the current $\underline{n}%
^{d}(t)$\ being chosen.)

Suppose $h(t_{e})=2$, i.e. the last two minority groups were '1' then '0'.
Hence $\Upsilon (t_{e})=0,$\ $h(t_{e}+1)=0$ and consequently

$\underline{S}(t_{e}+1)=\left( 
\begin{array}{c}
4 \\ 
-2 \\ 
-2 \\ 
0 \\ 
0 \\ 
2 \\ 
2 \\ 
-4
\end{array}
\right) .$

\bigskip

An expression for the time-averaged quantity called the `volatility'
(standard deviation of the number of agents choosing one particular group)
can be easily found using the above formalism: 
\begin{equation}
\sigma _{MG}=\frac{\bigg[\sum\limits_{t=t_{1}}^{t_{2}}\bigg[\varepsilon (t)-%
\overline{\varepsilon }\bigg]^{2}\bigg]^{\frac{1}{2}}}{t_{2}-t_{1}}
\end{equation}
where $\varepsilon (t)=[\underline{n}(t)^{T}\Sigma ]_{h(t)+1}\;$and$\;%
\overline{\varepsilon }$ is the time-average of $\varepsilon (t)$ from time $%
t_{1}$\ to $t_{2}$. Here $t_{1}$ and $t_{2}$\ denote the time window over
which the volatility is calculated. In the reduced strategy space \cite
{challet} a similar quantity to this standard deviation can also be written
down using our previously introduced (time-averaged) crowd-anticrowd
framework \cite{us2}: 
\begin{equation}
\sigma _{CA}=\frac{\sum\limits_{t=t_{1}}^{t_{2}}\bigg[\frac{1}{4}%
\sum\limits_{r=1}^{2^{m}}[\underline{n}(t)_{r}-\underline{n}%
(t)_{2^{m+1}+1-r}]^{2}\bigg]^{\frac{1}{2}}}{t_{2}-t_{1}}\ .
\end{equation}
For a given run of the game $\sigma _{MG}\neq \sigma _{CA}$, however these
quantities become quantitatively the same (within the limits of sample size)
when averaged over initial configurations of strategies\cite{us2}. $\sigma
_{CA}$ mirrors the semi-analytic approach introduced to motivate the
time-independent crowd-anticrowd theory of Ref. \cite{us2} (see Fig. 1 of
Ref. \cite{us2}). Indeed, the dynamical equations can be linked more
formally with our previous time-averaged approach\cite{us2}. Consider, for
example, the situation where no two strategies are tied in virtual points
and there are an equal number of agents having each possible pairing of
strategies (low $m$ limit and reduced strategy space), i.e. all elements in $%
\Omega $\ are equal and non-zero. It is then easy to show that $\underline{n}%
^{0}(t)_{r}$ reduces to $\underline{n}^{0}{}_{r}=\frac{N}{(2^{m+1})^{2}}[%
1+2(2^{m+1}-\rho (t)_{r})]$; this is precisely the vector of the quantity $%
n_{r}$ introduced in Ref. \cite{us2} now written in order of increasing
strategy label. If we allow for tied strategies, $\underline{n}^{d}(t)$\
will be non-zero thus reducing the size of large crowds and increasing the
size of the smaller crowds (and hence anticrowds), thereby leading to a
smaller standard deviation.

We now turn to a comparison between the standard deviation or `volatility' $%
\sigma$ obtained from numerical simulations and our (time-averaged)
crowd-anticrowd theory. We start with the basic MG. Figure 1 shows the
spread of numerical values for different numerical runs (open circles), the
full crowd-anticrowd theoretical calculation (large solid circles) and
various limiting analytic curves (solid lines) for which closed-form
expressions were given in Ref. \cite{us2}. Fuller details are provided in
Ref. \cite{us2}. The time-averaged dynamics can be described using a
quantity $P(r^{\prime }={\bar{r}})$ which represents the probability that
any strategy $r^{\prime }$ is the anti-correlated partner of strategy $r$ \ 
\cite{us2}. To produce the limiting analytic curves in Fig. 1, $P(r^{\prime
}={\bar{r}})$ is taken to be either a delta-function or a flat distribution.
The full theory takes the relevant form of $P(r^{\prime }={\bar{r}})$ from
the game. The agreement is very good, confirming that our theory captures
the essential physics.

In a variant of the basic MG, agents pick which strategy to use
stochastically at each timestep. Focussing on $s=2$, numerical simulations
\cite{sherrington} found that the larger-than-random $\sigma $ in the
`crowded' regime (i.e. small $m$) becomes smaller-than-random when the
strategy-picking rule is made increasingly stochastic. Our crowd-anticrowd
theory provides a quantitative explanation of this effect. Let $\theta $ be
the probability that the agent uses the worst of her $s=2$ strategies.
Figure 2 shows a comparison between numerical simulation (open circles) and
analytic expressions (monotonically-decreasing solid lines) obtained using
our crowd-anticrowd theory (full details are given in Ref. \cite{usstoch}).
These analytic expressions vary in their choice of $P(r^{\prime }={\bar{r}})$%
: the upper line $\sigma _{delta}$\ in Fig. 2 assumes a delta function while
the lower line $\sigma _{flat}$\ assumes a flat distribution. The theory
agrees well in the range $\theta =0\rightarrow 0.35$ and provides a
quantitative, yet physically intuitive, explanation for the previously
unexplained transition in $\sigma $ from larger-than-random to
smaller-than-random as $\theta $ increases.

Above $\theta =0.35$, the numerical data tend to flatten off while the
analytic expressions predict a decrease in $\sigma $ as $\theta \rightarrow
0.5$. This is because the analytic theory averages out the fluctuations in
strategy-use at each time-step. In Ref. \cite{usstoch} we showed how to
correct this shortcoming of the analytic theory. Consider $\theta =0.5$;
Fig. 2 inset (a) shows the measured numerical distribution in $\sigma $ for $%
\theta =0.5$, while inset (b) shows the result from the semi-analytic
procedure introduced in Ref. \cite{usstoch}. The two distributions are in
good agreement. Note that the non-zero average (4.7 for $N=101,m=2$ and $s=2$%
) for each distribution lies \emph{below} the random coin-toss limit $\sqrt{N%
}/2$. It is also possible to perform a fully analytic calculation of the
average $\sigma _{\theta }$\ in the $\theta \rightarrow 0.5$ limit \cite
{usstoch}; this value (which is also 4.7 for $N=101,m=2$ and $s=2$) is shown
in Fig. 2.

In summary, we have demonstrated that the crowd-anticrowd approach can be
applied to explain many aspects of MG games, yielding both time-averaged and
time-dependent theories (see also Ref. \cite{crowd}). Our efforts to develop
such simplified market games in order to describe real-world financial
markets are described elsewhere \cite{Paul}.

We thank A. Short for useful discussions.

\pagebreak

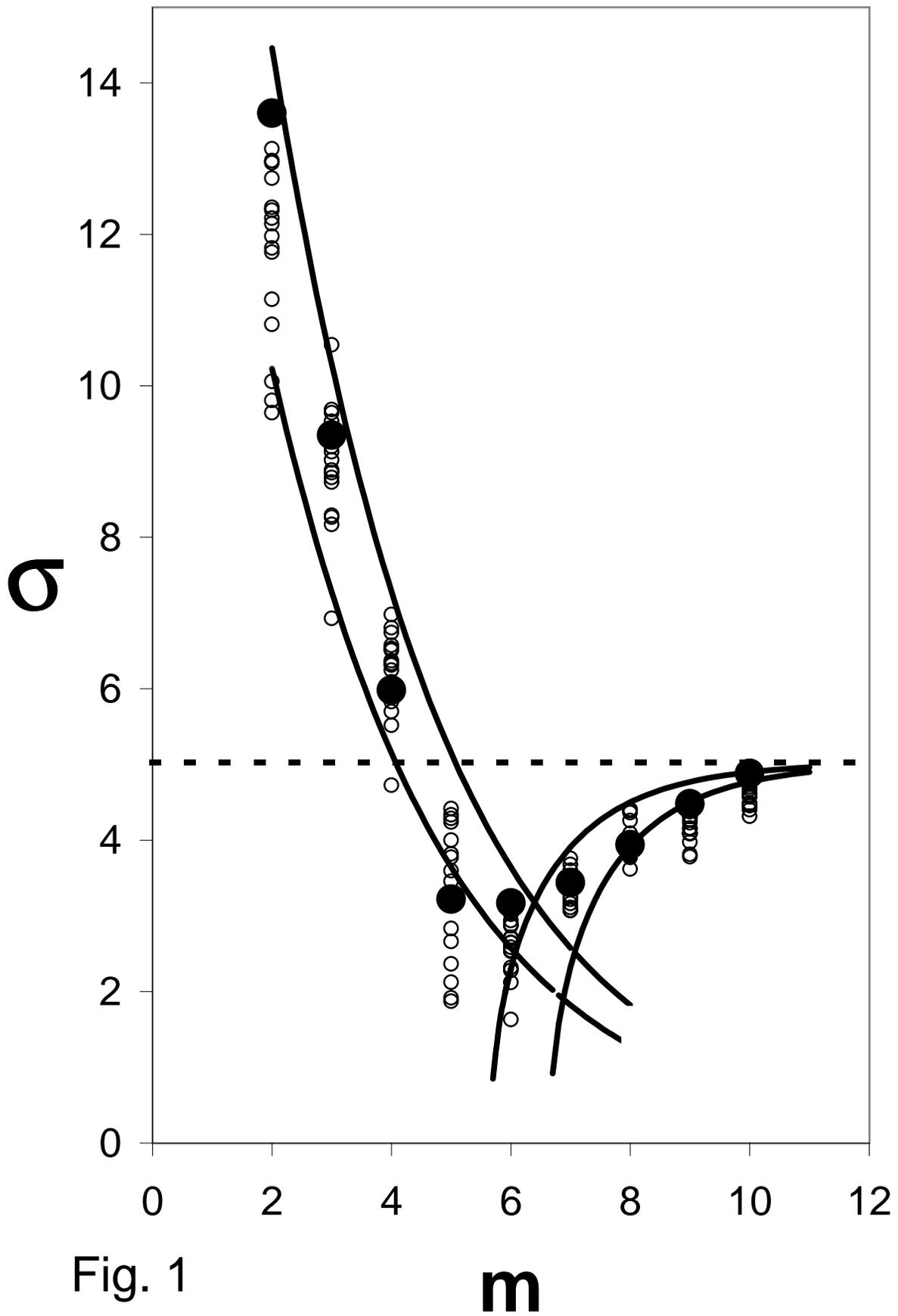
FIG. 1. Theoretical crowd-anticrowd calculation (solid circles) and
numerical simulations (open circles) for the standard deviation $\sigma$ in
basic MG with $s=2$ and $N=101$. 16 numerical runs are shown for each $m $.
Solid lines correspond to analytic expressions representing special cases
witihin the time-averaged crowd-anticrowd theory of Ref. \cite{us2}.

\bigskip

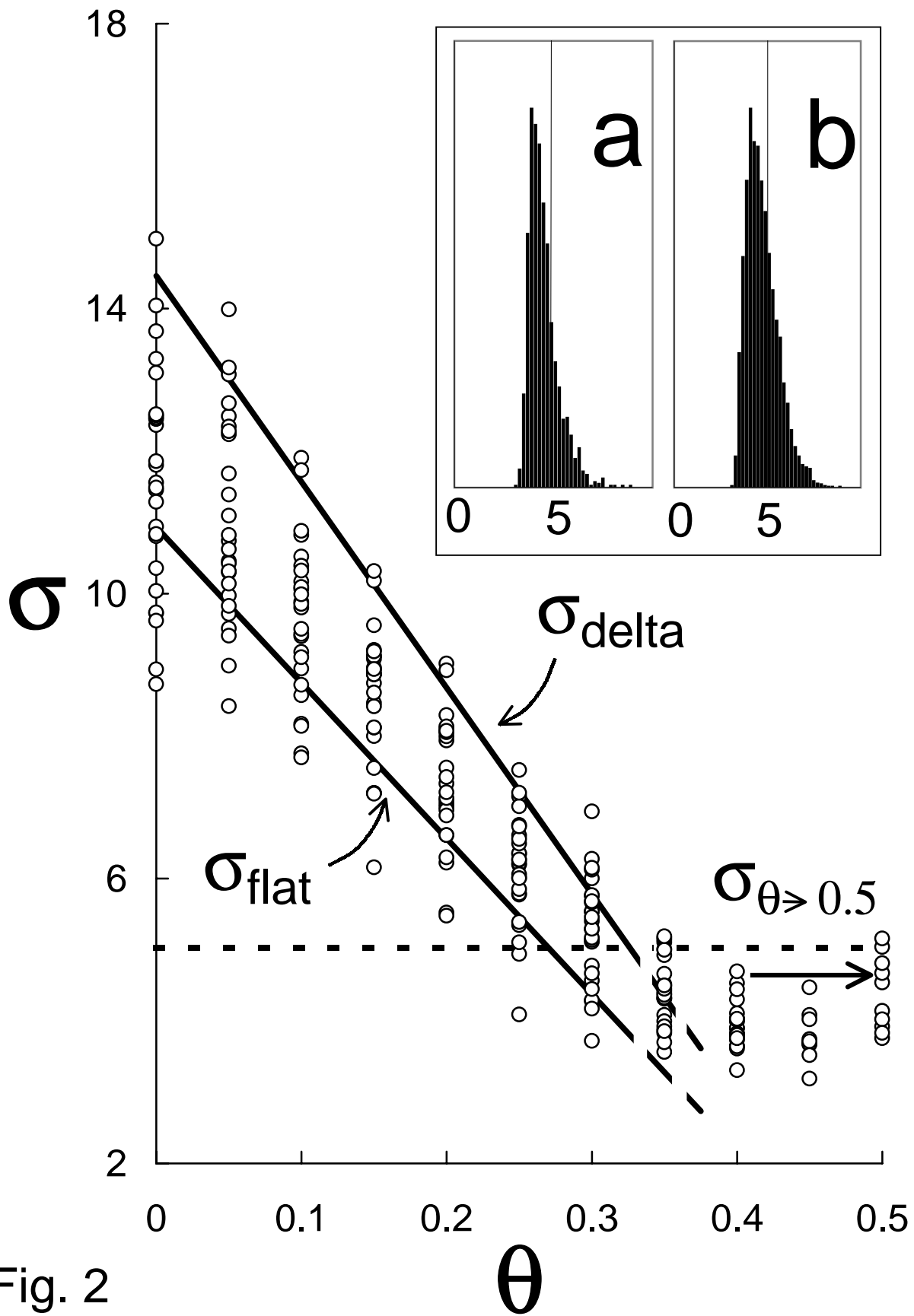
FIG. 2. Theoretical crowd-anticrowd calculation and numerical simulations
(circles) for $\sigma$ vs. the probability parameter $\theta$ in the
stochastic MG. Here $N=101,m=2$ and $s=2$. Monotonically decreasing solid
lines correspond to analytic expressions $\sigma _{delta}$ and $\sigma
_{flat}$ (see text). Dashed line shows random coin-toss value. Solid arrow
indicates theoretical value $\sigma _{\theta \rightarrow 0.5}=4.7$\ for $%
\theta \rightarrow 0.5$. Inset shows distribution of $\sigma $\ values at $%
\theta =0.5$\ for several thousand randomly-chosen initial strategy
configurations: (a) numerical simulation, (b) semi-analytic theory of Ref. 
\cite{usstoch}.

\end{document}